\documentclass[%
reprint,
superscriptaddress,
preprintnumbers,
nofootinbib,
amsmath,amssymb,
aps,
prd,
]{revtex4-1}

\usepackage{graphicx}
\usepackage{dcolumn}
\usepackage{bm}

\usepackage[utf8]{inputenc}
\usepackage[T1]{fontenc}
\usepackage{newtxtext,newtxmath}
\usepackage[dvipsnames]{xcolor}
\usepackage{color}
\usepackage{multirow}
\usepackage{threeparttable}
\usepackage{booktabs}
\usepackage{amsmath}
\usepackage{float}
\usepackage[colorlinks=true,pdfstartview=FitV,linkcolor=blue,citecolor=blue,urlcolor=blue]{hyperref}

\usepackage{enumitem}
\setlist[enumerate]{label*=\arabic*.}

\usepackage{xcolor}

\newcommand{\gap}{\mbox{$E_g$}}
\newcommand{\eh}{\mbox{$\epsilon_{eh}$}}
\newcommand{\F}{\mbox{$F$}}
\newcommand{\rec}{\mbox{$E_r$}}
\newcommand{\plasmon}{\mbox{$\hbar\omega_{pl}$}}
\newcommand{\optical}{\mbox{$\hbar\omega_0$}}
\newcommand{\qy}{\mbox{$\langle N(E) \rangle$}}

\begin{document}

{\small \hfill  FERMILAB-PUB-20-139-AD-E}

\title{Ionization Yield in Silicon for eV-Scale Electron-Recoil Processes}




\author{K.~Ramanathan}\email{ramanathan@uchicago.edu}\affiliation{Kavli Institute for Cosmological Physics, The University of Chicago, Chicago, IL, United States} 
\author{N.~Kurinsky}\email{kurinsky@fnal.gov}\affiliation{Fermi National Accelerator Laboratory, Batavia, IL 60510, USA}\affiliation{Kavli Institute for Cosmological Physics, The University of Chicago, Chicago, IL, United States}

\date{\today}

\begin{abstract}
The development of single charge resolving, macroscopic silicon detectors has opened a window into rare processes at the $\mathcal{O}$(eV) scale. In order to reconstruct the energy of a given event, or model the charge signal obtained for a given amount of energy absorbed by the electrons in a detector, an accurate charge yield model is needed. In this paper we review existing measurements of charge yield in silicon, focusing in particular on the region below 1~keV. We highlight a calibration gap between 12\textendash50~eV (referred to as the ``UV-gap") and employ a phenomenological model of impact ionization to explore the likely charge yield in this energy regime. Finally, we explore the impact of variations in this model on a test case, that of dark matter scattering off electrons, to illustrate the scientific impact of uncertainties in charge yield.
\end{abstract}

\maketitle

\section{Introduction}

Recent developments in silicon (Si) based particle detectors, including cryogenic calorimeters \cite{cdms:single,Hong_2020} and pixelated quantum charge detectors \cite{sensei,alex2017depfet}, have ushered in a new era of $\mathcal{O}$(eV) sensitivity to resolving the deposited energy of particles that traverse through them. These devices, capable of counting individual charge-pairs, not only have specific particle physics applications \textemdash\ such as in placing constraints on the existence of light MeV scale dark matter that recoils off electrons \cite{dm-electron,Agnese_2018,Abramoff_2019,alex2017depfet}, or in probing non-standard model neutrino interactions \cite{Harnik_2012,connie} \textemdash\ but broad astronomical applications such as in exoplanet searches \cite{spie:skipper}. 

Common to all of these use cases is the need to precisely identify the energy of the external particle. Generically, particle detectors work by measuring the deposited energy in an absorber material by one of three main avenues: charge production (ionization), photon production (e.g. scintillation) or collective excitations (phonons and plasmons), with further down-conversions intermixing these different production modes. In a semiconductor like Si \textemdash\ where ionization plays a dominant role above the \textit{band gap} \gap\ ($\sim$1~eV) \textemdash\ the deposited \textit{recoil energy} \rec\ is often inferred from counting the number of electron-hole pairs created, $n$, by way of the \textit{mean energy-per-pair} \eh. Due to the concurrent emission of phonons during the ionization process, \eh\ $>$ \gap\ and \eh\ is only reflective of the aggregate response of the material. The \textit{Fano factor}, \F, quantifies the dispersion of $n$ for a given $E$ and is sub-Poissonian (\F<1).  For deposits $\gg$20~eV in Si the statistical nature of this ionization process leads to asymptotic behaviour in the values for \eh\ and \F\ and thus provides for a simple statistical relationship between the expected energy and what was counted.

In this paper, we demonstrate that this relationship is not straightforward in the low-count regime. We show that systematics on the order of 50\% can arise when applying ionization models in scientific applications due to both the finite band-gap and complex features of the band-structure of crystalline Si which are not averaged out, and that the width of the hole band, rather than plasmon interactions, has a visible impact on the charge yield in the regime between 12\textendash 50~eV. Disentangling the effect of ionization is vital for correctly attributing the response of the detector to the physics of dark matter or some other unmeasured process, such as potential signals from elastic nuclear recoil, the Migdal effect\footnote{Ionization induced by the sudden shift of a recoiling nucleus} \cite{ibe}, or other collective effects \cite{kurinsky2020dark,kozaczuk2020plasmon}. 

The existing literature on ionization response is vast and often delves deeply into the condensed matter foundations of this topic which perhaps does not serve well a practitioner from the particle-physics community. As such, this paper is concerned with summarizing and providing for a simple phenomenological model, well supported by data, to allow experimental collaborations using Si detectors to provide results on an equal footing. We provide tables of computed probabilities $p_{n}(E)$, interpreted as the probability to ionize $n$ charge-pairs as a function of the deposited energy, for $E\leq50$~eV and a closed functional form for $E>50$~eV. This paper also serves as a blueprint for constructing a low-energy ionization model for similar semiconductor targets, which we leave for future work.

\section{Modeling Quantum Yield} \label{sec:modeling}

The process we attempt to model is energy redistribution from an initial hot carrier\footnote{A hot carrier is any charge with momentum much larger than that accessible thermally. For high-purity Si at and below room temperature, all charge pairs generated by particle interactions can be considered hot carriers.} to the electron and phonon system, in particular the ionization of subsequent electron-hole pairs by the initial carrier, known as impact ionization. All of the initial recoil energy is given to a single electron-hole pair. The number of total electron-hole pairs created after the cascade process is typically calculated as 
\begin{equation}
n = \frac{E_r}{\epsilon_{eh}(E_r)}
\end{equation}
where \eh\ has been shown to be constant in the high-energy limit (see Table~\ref{tab:measurements}). 

At low energy, we know that this formula breaks down. In a perfect lattice, any ionizing interaction below \gap\ is energetically inaccessible, so this equation is undefined ($\epsilon_{eh}\rightarrow \infty$). For energies between \gap\ and 2\gap, only one electron-hole pair is allowed by energy conservation, forcing a direct relationship between energy and \eh\ to ensure the mean is fixed. The uncertainty in this function therefore enters between $E_r=2E_{g}$ and the high energy limit $E_r >> E_{g}$. 

This allows us to summarize the goal of this work as fully characterizing the behavior first of {\eh}:
\begin{equation}
\epsilon_{eh}(E_r) = \left\{
\begin{array}{cr}
\infty & E_r < E_{g}\\
E_r & E_{g} \le E_r < 2E_{g} \\
\epsilon_{imp}(E_r) & E_r \ge 2E_{g} \\
\epsilon_{eh,\infty} & E_r \rightarrow \infty
\end{array}
\right.
\end{equation}
where $\epsilon_{imp}(E_r)$, the mean energy imparted by impact ionization, is the unknown function. 

This process also has a variance $\sigma^2(E_r)$, commonly related to the mean energy by the Fano factor \cite{Fano}:
\begin{equation}
F(E_r)=\frac{\sigma^2(E_r)}{n_{eh}(E_r)} = \frac{\sigma^2(E_r)\epsilon_{eh}(E_r)}{E_r}
\end{equation}
This factor, too, has an energy dependence, and in the high-energy limit some measurements have been made, but this parameter is far less well constrained (see Table~\ref{tab:measurements}). From energy conservation, $F=0$ below 2\gap\, and like the mean, has an asymptotic limit. The function is therefore
\begin{equation}
F(E_r) = \left\{
\begin{array}{cr}
0 & E_r <2 E_{g}\\
F_{imp}(E_r) & E_r \ge 2E_{g} \\
F_{\infty} & E_r \rightarrow \infty
\end{array}
\right.
\end{equation}

This two-component model has been repeatedly validated for energies >6~keV (see Table~\ref{tab:measurements}). A straightforward extension to lower energies, a placeholder often used in literature when discussing low energy phenomena, is to modify the piece-wise descriptions above to $\epsilon_{imp}=E_r$ and $F_{imp}=0$ for $E_{g} < E_r < \epsilon_{eh}$; we refer to this as the ``Simple Model" hereafter. Our goal here however is to explore a phenomenological model which stitches together the near-gap and high-energy limits based on available experimental measurements. This requires a framework for calculating $\epsilon_{imp}$ and $F_{imp}$. We begin by breaking down the calculation into constituent components, and then explore calculations made under different assumptions, as well as implications for Fano factor modeling.



Following Ref.~\cite{Wolf98}, we calculate the number of electron-hole pairs generated as 
\begin{equation}
    n(E_{r}) = 1 + \int_{E=0}^{E_r-E_{g}}dE P(E,E_{r})\langle N(E) \rangle
\end{equation}
where $P(E,E_{r})$ is the probability of the interaction occurring with a given energy distribution between the electron and hole, and \qy\ is the quantum yield, the average number of charges produced by impact ionization by a carrier with initial energy $E$ above gap \footnote{Note that $E_{r}$ is the total absorbed energy, measured from the top of the valence band, while $E$ is the energy above the bottom of the conduction band for electrons or below the top of the valence band for holes.}, assumed identical for electrons and holes. By definition, if the carriers do not impact ionize any additional electron-hole pairs, $n(E_{r})=1$.

\subsection{Initial Energy Distribution} \label{IED}

\begin{figure}[t]
	\centering
	\includegraphics[width=0.48\textwidth]{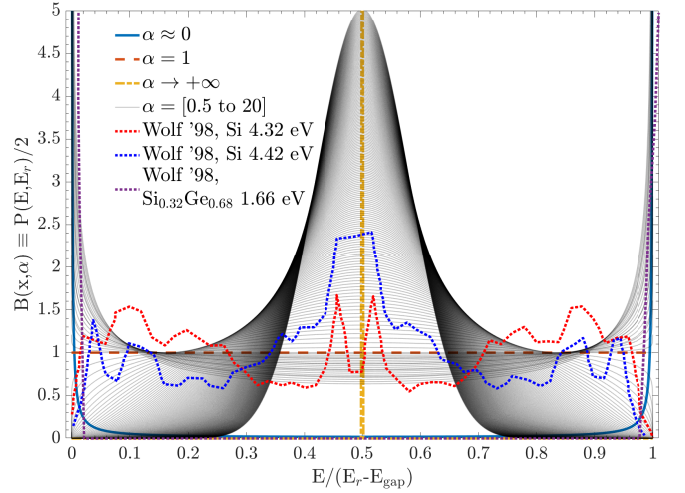}
	\caption{Evolution of the finite support Beta distribution, used to model the double probability distribution P(E,E$_r$), for selected values of shape parameter $\alpha$. The $\alpha \rightarrow 0$ case corresponds to all the deposited energy going to a single carrier. The $\alpha=1$ scenario is a uniform distribution, while $\alpha \to \infty$ corresponds to an equal energy splitting. Overlaid (dotted lines) is data extracted from Wolf \cite{Wolf98} for Si at 4.32 and 4.42 eV (derived from internal quantum efficiency measurements of a Si solar cell) and Si$_{0.32}$Ge$_{0.68}$ at 1.66 eV (numerical calculations), indicating the general evolution even over a small energy range between the different regimes of energy partitioning between hot carriers.} 
	\label{fig:betadistribution}
\end{figure}

The role of $P(E,E_{r})$ is to describe how energy is distributed between the carriers in the electron-hole pair, and is normalized in $E$ by definition to 2 (one electron and one hole per pair). The full treatment used by Ref~\cite{Wolf98} is to treat carriers equally, such that this function obeys the symmetry relationship
\begin{equation}
    P(E,E_r) = P((E_r-E_{g})-E,E_r)
    \label{eq:ProbSplit}
\end{equation}
where, for small energies, we find the distribution is highly peaked around $E=0$ and $E=E_r-E_{g}$. The robust calculation involves a summation over allowed energy states for the conduction and valence bands given the transition matrix element for photon absorption, and that model will be included in our comparison of calculations. The symmetry of this function, however, allows for three simplifying assumptions to capture the full range of possible outcomes:
\begin{enumerate}
\item $P(E,E_r)=\delta(E)+\delta(E_r-E_{g}-E)$ -- A peaked distribution for maximal energy imbalance, approximately true for very low energy transfer (referred to in this paper as `all to one');
\item $P(E,E_r)=2\left[E_r-E_{g}\right]^{-1}$ -- A flat distribution, approximately true for $E_r >> E_{g}$ (referred to in this paper as `uniform');
\item $P(E,E_r)=2\delta(E_r-E_{g}-2E)$ -- A peaked distribution at half of the above-gap energy, which occurs for $E_r$ around resonance features near the $\sim3.4$\textendash$4.2~eV$ direct-gap transitions in Si (referred to in this paper as `equal split'). This is also the case that minimizes overall impact ionization, which is a strong function of energy.
\end{enumerate}

We extend these three cases to the one-component model described by the one-parameter beta distribution
\begin{equation}
    P(x|\alpha) = \frac{2}{B(\alpha)}x^{\alpha-1}(1-x)^{\alpha-1}
\end{equation}
where $x=\frac{E}{E_r-E_{g}}$ and $B(\alpha)$ is the one-parameter Beta function used to normalize the probability distribution. One can see that this function has the same symmetry as $P(E,E_r)$, and each case above has a corresponding $\alpha$ value: case 1 corresponds to $\alpha\rightarrow 0$, case 2 to $\alpha=1$, and case 3 to $\alpha\rightarrow \infty$. These cases, and more general cases for a range of $\alpha$ values found in this paper, are shown in Fig.~\ref{fig:betadistribution}.

We can qualitatively compare the shape of this distribution to the calculations done by Ref.~\cite{Wolf98}, to set expectations for how $\alpha$ scales with energy. We see that, for $E_r\sim E_{g}$, the excess energy is given entirely to either the electron or hole, and we expect $\alpha \rightarrow 0$ in the low energy limit. Around $E_r=4.3-4.4$~eV, see Fig.~\ref{fig:betadistribution}, we observe a transition from equal energy split to more uniform energy sharing, corresponding to a rapid increase in $\alpha$ through 1 to $\alpha > 1$. Ref.~\cite{scholze1996} notes that the hole valence band width $W$ is only 12~eV wide, and thus we expect in the high-energy limit that the electron takes the majority of the energy; so we expect the applicability of our $P(E,E_{r})$ description to lessen due to its inability to capture the narrower allowed space for the hole energy and the asymmetry of the distribution.

\subsection{Impact Ionization Model} \label{subsec:IIM}

The second component of the yield model is the impact ionization function \qy, which we recollect describes the mean number of electron-hole pairs produced by a hot carrier with initial energy $E$. This function is bounded by two extremes; in the limit of maximal impact ionization, up to \qy$=E/E_{g}$, electron-hole pairs can be created (rounding down to the nearest integer), and in the limit of no impact ionization, \qy$=0$. In the second case, energy is largely dissipated by phonon emission, meaning that the true \qy\ is thus determined by a rate balance between impact ionization and phonon emission as a function of energy. Ref.~\cite{Jacoboni} shows that the dominant phonon-scattering mechanism in these energy ranges, both by rate and total energy dissipated, is through the emission of optical phonons, and thus acoustic phonon emission can be neglected in impact ionization modeling.

As in Ref.~\cite{Wolf98}, we adopt the impact ionization model of \citet{ABS}. In this model, only two energy dissipation processes are considered: electron-hole pair creation, and emission of a phonon of energy \optical. In Si, \optical$\sim$63~meV \cite{Jacoboni}, so a charge carrier above gap can easily create many optical phonons. If the rate of electron-hole pair creation is $\Gamma_{eh}(E)$ and phonon production is $\Gamma_{ph}(E)$, then the probability of generating an electron-hole pair at a given energy is dependent only on the ratio of these rates, found to be
\begin{equation}
    \frac{\Gamma_{ph}(E)}{\Gamma_{eh}} = A\frac{105}{2\pi}\frac{\left(E-\hbar\omega_0\right)^{1/2}}{\left(E-E_g\right)^{7/2}}.
    \label{eq:ratio}
\end{equation}
Here A is a constant of the system, defined as
\begin{equation}
    A = \frac{|M_{ph}|}{|M_{eh}|}\frac{4\pi^4}{V\Delta}\left(\frac{\hbar^2}{2m}\right)^3
    \label{eq:Aconstant}
\end{equation}
where $V$ is the semiconductor volume, $\Delta$ is the volume per electronic state, $m$ is the free particle mass, and $|M_{ph}|$ ($|M_{eh}|$) is the phonon (electron) scattering matrix element. This enables us to calculate the charge production probability of a particle with energy $E_i$, using Eq.~\ref{eq:ratio}, as
\begin{equation}\label{eq:pn}
    p_{eh} = \left[1+\frac{\Gamma_{ph}(E_i)}{\Gamma_{eh}(E_i)}\right]^{-1}.
\end{equation}
The elegance of this model is that it is able to reduce the complex micro-physics of the problem to one phenomenological constant, $A$, which can be tuned to match experimental values. This is beneficial due to the complex nature of electron-electron interactions at this energy scale.

There are a number of simplifying assumptions made in this model which need to be explicitly stated. The electrons and holes are assumed to be free particles to first approximation, and therefore scattering is isotropic and effective masses are vacuum masses. It assumes all states are equally accessible, and therefore that the matrix element for each transition is identical. We do however constrain the hole energy $E_h \leq W$. All these assumptions are akin to assuming interactions are highly athermal and occur far enough above the band-gap that the detailed band structure is negligible. It also simplifies phonon scattering substantially, allowing for a single, quantized phonon energy, ignoring the multiple optical phonons and the continuum of acoustic phonon energies \cite{Jacoboni}. 

The latter assumption is justified by the rate difference mentioned earlier; the optical phonons all have comparable energy, so the impact of having multiple distinct energies is small. The former assumption comes from the high density of states for particles far above the band-gap, but nonetheless makes $A$ a non-physical parameter, and requires explicit validation of this model before it can be considered predictive. The benefit is that $A$ can be tuned to produce the correct \eh\ in the large energy limit; comparing this to the other experimental data, given only a single degree of freedom, the success of the model matching data across many energy scales validates the assumptions that have gone into it (see e.g. Refs.~\cite{chang,Wolf98,cartier}). Nevertheless, we should keep in mind that, in practice, $A$ may be a function of temperature or substrate purity, among other possible effects. This is likely to be the leading systematic in applying this model to regimes which have not been validated (e.g. $T<5$~K, or highly-doped substrates).

The last mode of energy dissipation that needs to be accounted for is plasmon production\cite{Pines}. The model of Ref.~\cite{ABS} makes the simplifying assumption that energy redistribution is largely done through conservative creation of plasmons above the plasmon energy \plasmon, and below this energy energy redistribution proceeds according to the charge-phonon scattering balance model. This implies that charge production should be linear above this energy, which matches what is measured experimentally (see Table~\ref{tab:measurements}). In our implementation, we split the total energy into $n_{pl}=\lfloor(E_r-E_{g})/\hbar\omega_{pl}\rfloor$ plasmons of energy \plasmon\, with a final electron that has energy $E_r-E_{g}-n_{pl}\hbar\omega_{pl}$. This implicitly assumes that the total energy is evenly divided among plasmons of a fixed energy, and that those plasmons decay by production of an electron-hole pair with total energy equal to the plasmon energy. Both of these assumption are non-physical, but we do not observe any impact on the ionization yield by adding or removing this mechanism in agreement with the conclusions of Ref.~\cite{ABS}. We include plasmons in this way to be consistent with the model described by Ref.~\cite{ABS}, but will discuss the impact of plasmons, if any, on the ionization yield in Section~\ref{sec:discussion}.

\subsection{Temperature Dependence} \label{subsec:temp}

\begin{figure}[t]
	\centering
	\includegraphics[width=0.48\textwidth]{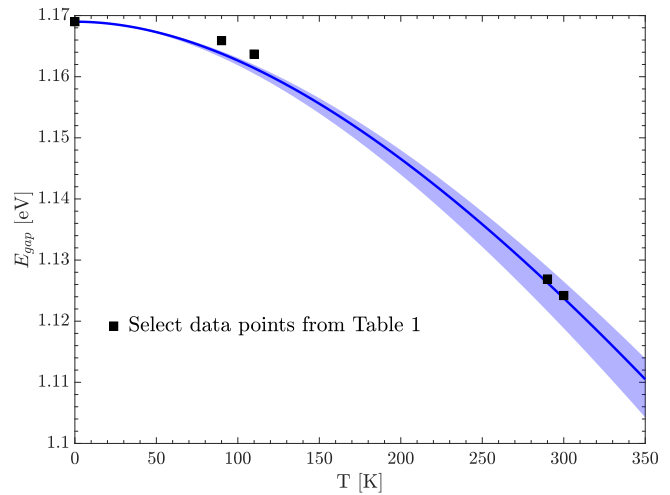}
	\caption{Evolution of band-gap as a function of silicon temperature from Eq. \ref{eq:varshini} (solid line) showing $\pm 1 \sigma$ bands (shaded region) and a characteristic sample of data points from Table \ref{tab:measurements}.}
	\label{fig:varshini}
\end{figure}

A final, important consideration we make in this paper is the effect of temperature dependence on the ionization yield. Detectors relevant to rare event searches are often operated cryogenically in order to mitigate high dark rates at room temperature, in the temperature range below 120~K down to $\sim$10~mK. Given the simplified nature of the band structure for the ionization model discussed above, we incorporate temperature dependence purely through the variation of the gap energy, given by Eq.~\cite{varshni} :
\begin{equation}
E_g(T) = E_g(0) - \frac{aT^2}{T+b}
\label{eq:varshini}
\end{equation}
where we take E$_g$(0)=1.1692, $a=(4.9\pm0.2)\times 10^{-4}$~eV and $b=655\pm 40$~K following the results of Ref.~\cite{alex96} who experimentally measure the photoluminescence spectra of crystalline silicon up to 1000~K. The resulting gap energies, plotted in Fig.~\ref{fig:varshini}, are in general agreement with other values in Table~\ref{tab:measurements}. It should be noted that this equation is purely phenomenological, but can successfully fit the temperature dependence of the band-gap across many semiconductors, as first indicated in Ref.~\cite{varshni}. Other forms, such as piece-wise quadratic fits \cite{bludau}, have been used as well, and give similar results. 

While one can discuss tuning the gap energy independently without modifying temperature, that is a distinct effect with many potential causes. We recognize the primacy of temperature as an experimental input, and thus argue that any change in $T$ can be seen to first approximation as a change in \gap.

The only other parameters that may show a temperature dependence are $A$ and the phonon energy $\hbar\omega_0$. Because these processes are highly athermal, however, we fix them in our model, and expect that any change in these parameters due to temperature is subdominant to the systematic uncertainties on their nominal values. We will show in Section~\ref{sec:results} that only the change in gap is necessary to match temperature dependent results from the literature, validating this approach.

\section{Data}

\begin{table*}[htp]
    \centering
    \begin{threeparttable}
    \caption{Literature values for the Fano factor $F$, mean energy per electron-hole pair \eh\ in the high-energy limit, band-gap energy \gap\, the ratio $A$ of phonon-carrier to carrier-carrier scattering, the optical phonon energies $\hbar\omega_o$, and the plasmon energy $\hbar\omega_pl$. An earlier version of this table specific to Fano factor can also be found in \cite{fraser1994}. We summarize the energy gap at a few key temperatures, but all references have many more data points and focus on fitting measurements to the functional form of \cite{varshni}; see there for more details. We exclude band-gap data from \cite{MacFarlane}, fit in \cite{varshni}, due to discrepancy with more modern methods which have seen widespread adoption (see for example \cite{Canali72}). For the last three values, these are not from quantum yield measurements. The \cite{chang} values come from hot-electron injection measurements. The other two values come from the sources specified above.}
    \label{tab:measurements}
    \begin{tabular}{|c|c|c|c|c|}
    \hline
        Parameter & Value & Temperature & Source & Reference \\
        \hline
        \multirow{4}{*}{F} & 0.118 & 110 -- 240~K & 5.9~keV $\gamma$ & \cite{lowe} \\
         & 0.117 & 180~K & 5.9~keV $\gamma$ & \cite{mccarthy1995} \\
         & 0.14 -- 0.16 & 180~K & 2 -- 3.7~keV $\gamma$ & \cite{owens}\tnote{a} \\
         & 0.128 & 130~K & 5 -- 8 ~keV $\gamma$ & \cite{rama} \\
         & 0.119 & 123~K & 5.9~keV $\gamma$ &\cite{rodrigues2020} \\
        \hline
         & 3.66~eV & 300~K & 1~eV -- 1~keV $\gamma$ & \cite{scholz2000} \\
         & 3.66~eV & 300~K & 115 -- 136~keV e,$\gamma$ & \cite{pehl} \\
         & 3.63~eV & 300~K & 1~MeV $e^{-}$, 5.5~MeV $\alpha$ & \cite{ryan} \\
         & 3.62~eV & 300~K & 5.5 -- 6.3~MeV $\alpha$ & \cite{pehl} \\
    \eh\  & 3.67~eV & 180~K & 2 -- 3.7~keV $\gamma$ & \cite{owens} \\
         & 3.749~eV & 123~K & 5.9~keV $\gamma$ & \cite{rodrigues2020} \\
         & 3.75~eV & 110~K & 5.9~keV $\gamma$ & \cite{lowe} \\
         & 3.70~eV & 100~K & 5.5~MeV $\alpha$ & \cite{Canali72} \\
         & 3.72~eV & 6 -- 70~K & 480~keV $\gamma$ & \cite{Dodge} \\
         & 3.72~eV & 5~K & 5.5~MeV $\alpha$ & \cite{Canali72} \\
         \hline
         \multirow{6}{*}{\gap} & $\sim$1.12 & 300~K & \multirow{6}{*}{Photoabsorption} & \cite{alex96} \\
         & 1.127 & 290~K & & \cite{bludau} \\
         & 1.164 & 110~K & & \cite{bludau} \\
         & 1.166 & 90~K & & \cite{bludau} \\
         & 1.169 & 0~K & & \cite{alex96} \\
         & 1.170 & 0~K & & \cite{bludau} \\
         \hline
         A & 5.2~eV$^2$\tnote{b} & 300~K & 2 -- 5~eV $e^{-}$ & \cite{chang} \\
         \hline
         $\hbar \omega_0$ & 59~meV (TO), 62~meV (LO) & N/A & DFT\tnote{c} & \cite{Jacoboni} \\
         \hline
         $\hbar\omega_pl$\tnote{d} & 16.6 $\pm$ 0.1 ~eV & N/A & EELS\tnote{e} & \cite{Chen} \\
    \hline
    \end{tabular}
    \begin{tablenotes}
        \item[a]{See also \cite{owens1996}}
        \item[b]{Data compared to the value obtained by \cite{ABS}}
        \item[c]{Calculated from density functional theory (DFT), assumed temperature independent; see \cite{Jacoboni} for more details.}
        \item[d]{We did not do an exhaustive survey of plasmon energy measurements as they were not important for the detailed low-energy modeling, but we expect there is some uncertainty in this value beyond the statistical uncertainty on this measurement.}
        \item[e]{Electron energy-loss spectroscopy}
    \end{tablenotes}
    \end{threeparttable}
\end{table*}

The data for quantum yield and Fano factor in Si considered in this paper are summarized in Table~\ref{tab:measurements}. Most of the available measurements of \F\ and \eh\ are made at high energy, and these measurements are broadly consistent with each other. Of the data in Table~\ref{tab:measurements}, there are only a few references which make measurements between 2.4~eV and 1~keV, the energy range in which our models show the most variation:
\begin{enumerate}
    \item \citet{chang} measure electron impact ionization via injection of hot electrons of known energy into a Si transistor. Their measurements validate the impact ionization model of \citet{ABS} up to 5~eV.
    \item \citet{Wolf98} measure quantum yield for photon absorption between 2.5~eV and 5~eV, and compare it to the quantum yield predicted by the \citet{ABS} model and an energy-sharing distribution determined from summing over momentum eigenstates from band structure calculations. They show that both aspects of the model are necessary to accurately reproduce measured quantum yield.
    \item \citet{scholz2000} measure of \eh\ at 300~K between 3 and 1500~eV, with a gap between 8~eV and 50~eV,using a Si photodiode in an X-ray beamline. We use these measurements to extract an expected curve for $\alpha$ as a function of photon energy at room temperature. These data are shown in Fig.~\ref{fig:ehfit}.
\end{enumerate}
The gap in the Ref~\cite{scholz2000} data is reflective of a broader ``UV-gap" in the region between VUV and X-ray energies caused both by lack of tunable sources and the very short mean free path of photons in this energy range in all materials (see e.g. Ref.~\cite{struder}). For wavelengths below $\sim$8~eV, photons from a thermal or athermal source can penetrate through thin windows and coatings, and enough deposition occurs in the Si to be distinguished from quenched surface events. Above 8~eV, very few table-top sources exist, and only specialized windows can transmit light with adequate efficiencies. At 50~eV and above X-ray fluorescence sources become available\footnote{The lowest K$\alpha$ line is found in Lithium at 52~eV}. At these energies, the photoelectric cross section also begins to drop, and high intensity light can be generated and propagated to the detector through thin metal windows \cite{scholz2000}. For these reasons very little data exist over this 40~eV energy gap, and in the following section we discuss the extrapolations we employ to stitch together the quantum yield across this gap.

\subsection{Monte-Carlo Simulation}

To compute $p_n$, \eh\, \F, and \qy\ we employ the Monte-Carlo algorithm outlined in Ref. \cite{alig83}, following the schematic shown in Fig. 1 of Ref. \cite{ABS}. A single external particle deposits energy E$_r$, and with selected parameters $A$, \optical\, and \gap\, triggers a cascade briefly outlined as follows:

\begin{figure*}[t!]
	\centering
	\includegraphics[width=0.99\textwidth]{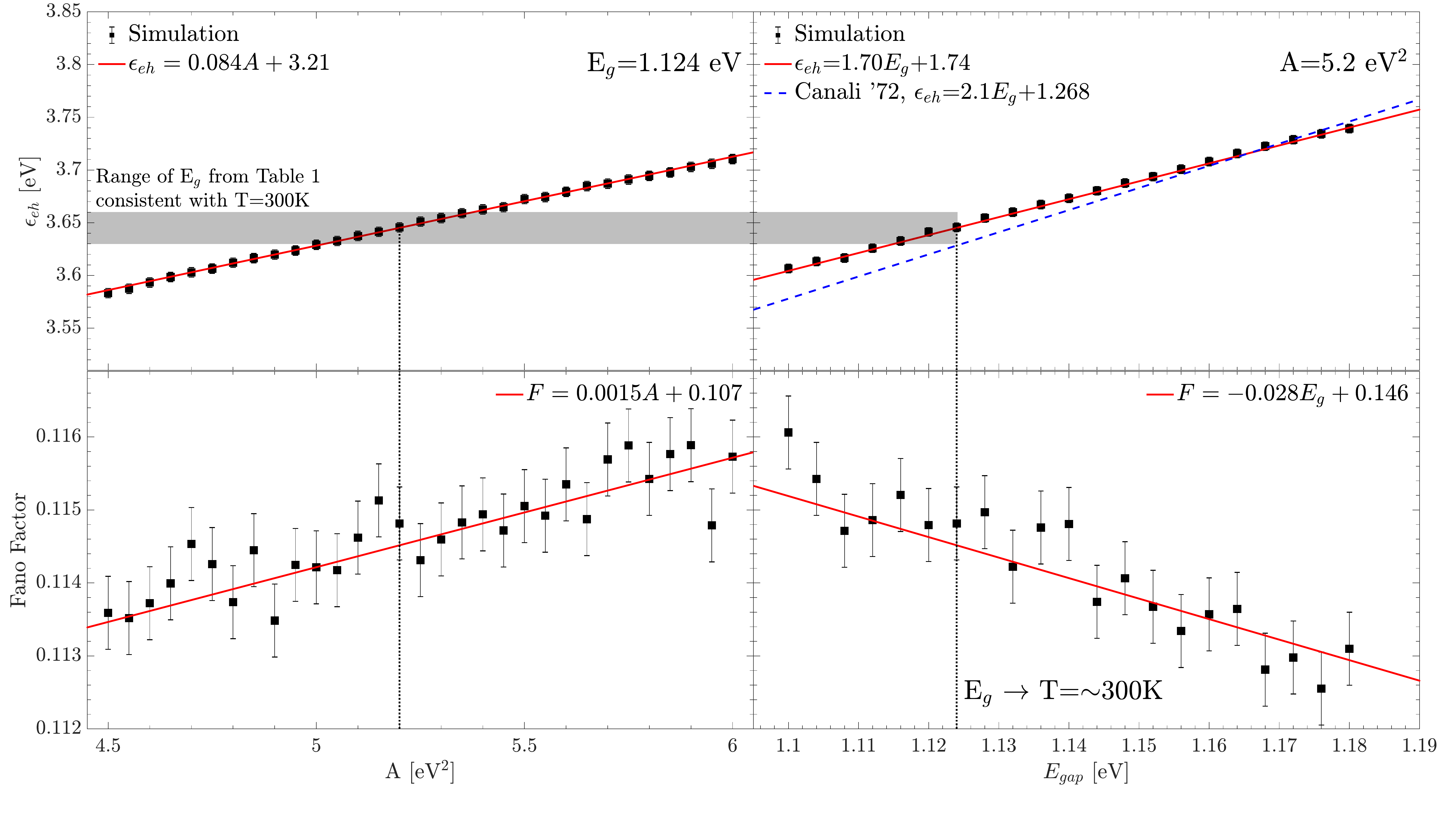}
	\caption{Left: Simulation of dependence of both mean energy-per-pair \eh\ (top) and Fano factor \F\ (bottom) at the 300~K slice (\gap = 1.124) on the $A$ parameter, the propensity for energy loss to occur due to phonon emission (larger $A$) versus ionization, as represented from Eq. \ref{eq:Aconstant}. Linear fits, and corresponding equations, are also provided (solid red lines). The scatter in data points is due to Monte-Carlo statistics. Right: Simulation of Mean energy-per-pair \eh\ (top) and Fano factor \F\ (bottom) dependence on band-gap energy \gap at the A=5.2 eV$^2$ slice. The temperature effect is expressed by varying the band-gap energy according to Eq. \ref{eq:varshini}. Linear fits, and corresponding equations, are also provided (solid red lines). A fit from Ref. \cite{Canali72} for \gap\ is also shown (dashed blue) demonstrating the general capture of features seen in data. The scatter in data points is again due to Monte-Carlo statistics.}
	\label{fig:AEGdependence}
\end{figure*}

\begin{enumerate}
    \item If $E_r>E_g$, we generate an electron and a hole with energies given by $P(E,E_r)$, but with the imposed constraint $E_h<W$; otherwise, the chain terminates.
    \item \label{step1} We follow each particle as it down-converts. Any electrons with energy $>$\plasmon\ are assumed to create plasmons of quantity $n_{p}= \lfloor E_i / \hbar\omega_{pl} \rfloor$. These are individually treated as impact ionizations.
    \item We calculate the charge production probability $p_n$ of a particle with energy $E_i$ using Eq.~\ref{eq:pn}. We select ionization or phonon production according to this probability.
    \item If a phonon is produced, then energy \optical\ is lost in the medium and the process loops back to step \ref{step1} with new energy $E_{i+1}=E_i-$\optical.
    \item If instead an ionization event occurs then 3 new particles are effectively created \textemdash\ the original plus and electron and a hole, with a total energy of $E_i-E_g$ due to the release of the new electron-hole pair. We assume here that the conduction and valence bands are isotropic and parabolic, and that all states are available to the new scattered products. The split between these 3 particles results in energies $E_{i+1,e,h}$ with values given by integrating over the density of states (see Section II.B. in Ref. \cite{alig83}) and where $E_{i+1}$ is the new energy of the original particle. Here is where, post-hoc, we set $E_h \leq W$, also applicable to the original particle if it is a hole, and re-split the difference in energies uniformly between remaining carriers.
    \item If $E_{i+1}<$\gap then the process terminates and only $E_{e,h}$ are fed back into step \ref{step1} otherwise all 3 particles are independently looped back to step \ref{step1}.
\end{enumerate}

This process continues until all tracked particles have kinetic energy below \gap, including those produced by the plasmons, resulting in the production of $n$ electrons (and holes). Repeating this nested approach yields a distribution of charge pairs, normalization of which gives the requisite probability of of pair-creation $p_n$.

\section{Results}\label{sec:results}

We begin this section by exploring, through simulation, the effects of the parameters $A$ and \gap\ at $\sim$100~eV ($>>$\gap) on \eh\ and \F, allowing us to fold in the effects of temperature. Next, we present the results of fitting our single-parameter model to the data from Ref.~\cite{scholz2000}, to finally produce pair-creation probabilities $p_n(T)$. 

\subsection{Micro-physics \& Temperature dependence}

To investigate the effects of both temperature, by proxy of gap energy \gap, and changes in the probability of phonon emission $A$, we compute the dependence of both \gap and $F$ on these parameters via simulation. Based on linear behavior across both dimensions for both quantities, we identify the global relations 
\begin{align}
    \epsilon_{eh}&=1.7E_g + 0.084A + 1.3, \\
    F&=-0.028E_g + 0.0015A + 0.14
\end{align} 
via least-squares regression. We confirm the consistency of both the model and of selecting $A=5.2$~eV$^2$, matching the original derivation in Ref. \cite{ABS} and the empirical validation in Ref. \cite{chang}, by noting that the resultant \eh\ values are in agreement with Table \ref{tab:measurements} at 300~K, seen by the confluence of dashed lines in the planar slices of Fig. \ref{fig:AEGdependence}. 

Fano values, shown in Fig. \ref{fig:AEGdependence} (bottom) are constant at 2\% level, but undershoot literature as per Table \ref{tab:measurements}. Unlike for \eh\, this model is not tuned for a specific Fano factor, and is thus predictive. The discrepancy observed between Fano measurements, and between the model and measurements, can potentially be attributed to one-sided systematics inherent to the measurements we quote that serve to inflate the measured Fano factor (see Appendix~\ref{app:fano}). We note, however, that the most recent measurements of the Fano Factor in Si in Ref. \cite{rodrigues2020}, using a device with single charge resolution, are closer to our asymptotic value of $F_{\infty}=0.115$ than prior considerations. 

Finally, \eh\ tracks the relationship from the experimental setup of Ref. \cite{Canali72}, to within 0.5\%, allowing use to conclude that our single-parameter model, regardless of energy partition, is capable of reproducing measured \eh\ and \F\ for high-energy energy depositions.



\subsection{Energy Dependence}\label{sec:energy}

\begin{figure*}[t]
	\centering
	\includegraphics[width=0.48\textwidth]{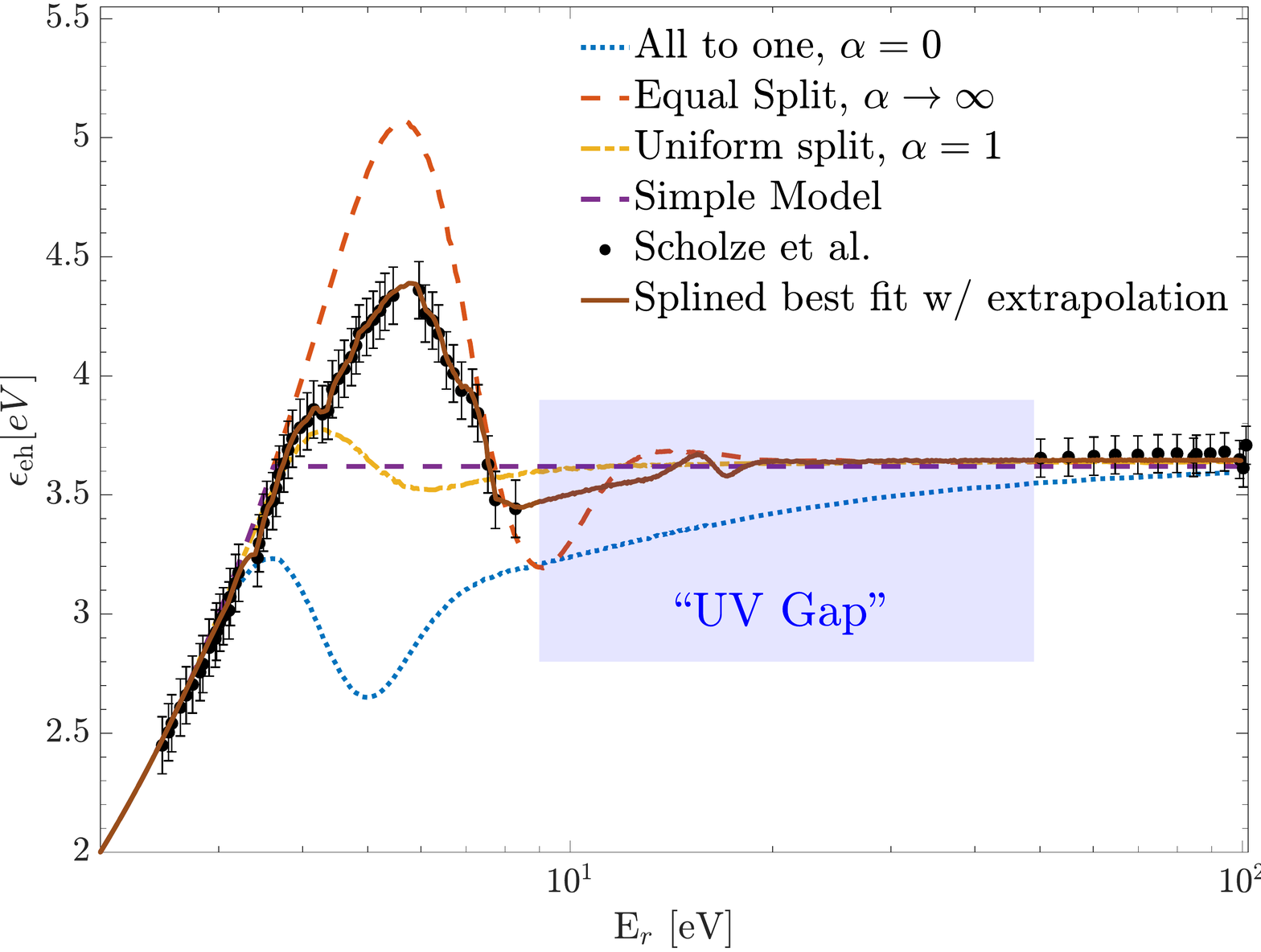}
	\includegraphics[width=0.48\textwidth]{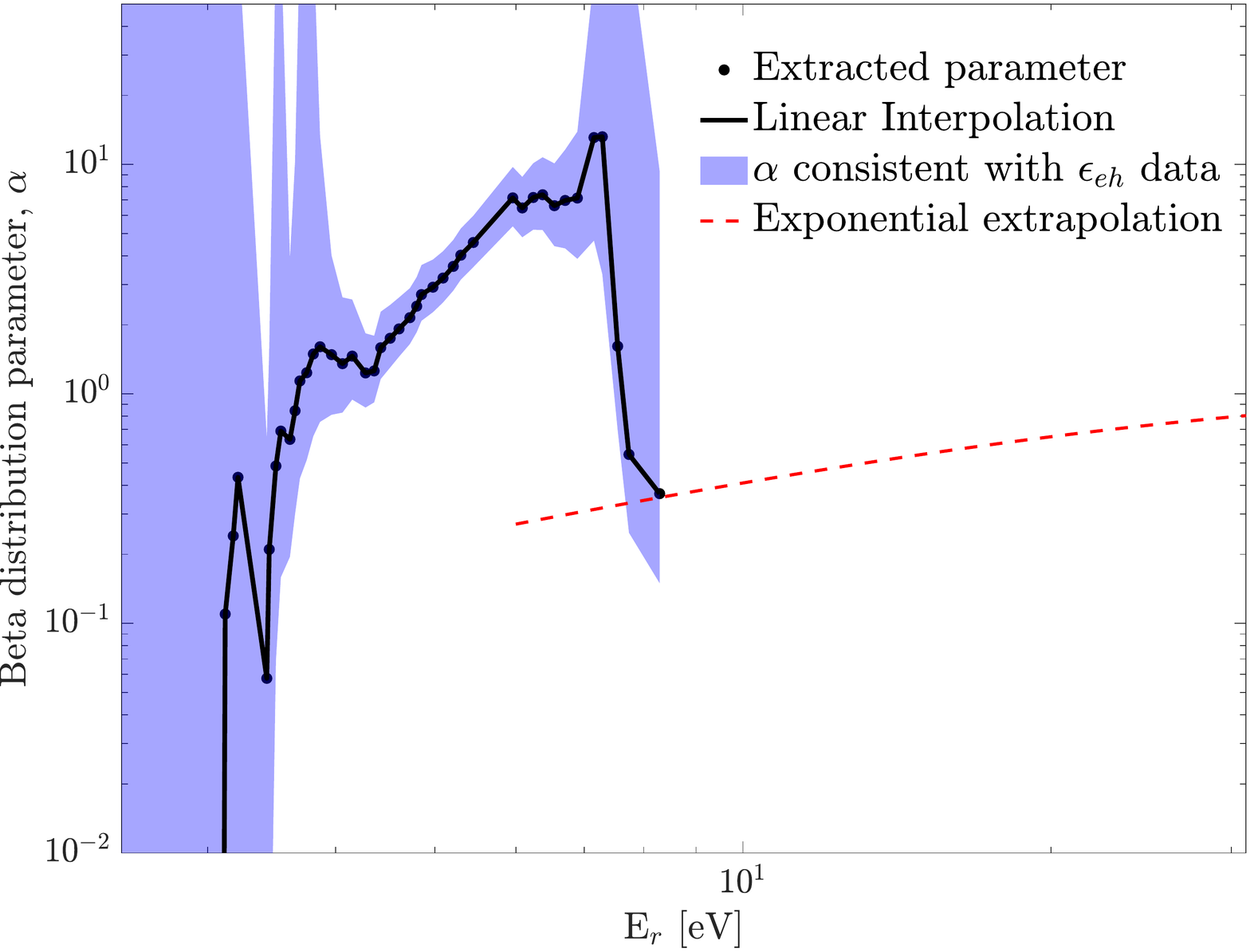}
	\caption{Left: Pair creation energy \eh\ for various energy partitioning schemes outlined in Section \ref{sec:modeling} and \ref{sec:energy}. Low-energy data points (black circles) \cite{scholz2000} between 2\textendash8~eV are fit using the model prescribed in this paper to give $\alpha$ values which are then interpolated and simulated back into \eh\ space (brown solid curve). The feature seen at 15\textendash 20~eV is a result of the imposed finite valence band width. Right: The extracted Beta distribution parameter, $\alpha$, with linear interpolation between neighboring points and single parameter exponential extrapolation in ``UV-gap" region as discussed in the text. The shaded bands represents the resultant fit on the $\pm 1 \sigma$ from the extracted Scholze points. However, some values of \eh\ are not recoverable, to within a tolerance of $1\times10^{-4}$, regardless of the $\alpha$ parameter, hence the shaded area encompasses $\alpha=[0,\infty]$ for certain energies.}
	\label{fig:ehfit}
\end{figure*}

We account for $P(E,E_r)$ by extracting it from data, specifically Ref. \cite{scholz2000}, by fitting to measured pair creation energy below 100~eV as a function of energy. This fit is performed assuming $A=5.2$~eV$^2$ and by setting $T=300$~K, the temperature at which these data were acquired. The left panel of Fig.~\ref{fig:ehfit} shows the mean energy-per-pair \eh\ as a function of initial energy for the 3 simplified energy distribution scenarios discussed in Section \ref{IED}. We note that the assumptions lead to the same behavior below 3~eV, and converge to the same value by $\sim$100~eV, but are largely discrepant in the energies between these points. None of the simplified models accurately reproduce the measured behavior between 3 and 10~eV; by 20~eV, all but the extreme $\alpha=0$ energy distributions have converged. We turn off the effects of $W$ and plasmon production when discussing the simplified energy distribution scenarios to more precisely disentangle their effects on the overall charge yield.

The lack of experimental data in the region between $\sim$9\textendash50~eV, often termed the ``UV-gap", necessitates the use of an extrapolation, where we have chosen to drive $\alpha$ to 1 parsimoniously using a single parameter exponential tied to the location of the last point. In and above this region, we expect a transition to the micro-physics based model of capping the maximum amount of energy transferable to a hole ($W$)\cite{scholze1996}. 

The right panel of Fig.~\ref{fig:ehfit} shows the best-fit values for $\alpha$ in the range between 0 and 8~eV, and the left panel demonstrates that the mean energy inferred from each fit point is an excellent fit to the data. The structure observed implies the following:
\begin{enumerate}
    \item Up to $\sim$4~eV, the data is consistent with a uniform energy distribution; there is little enough impact ionization, however, that we can only really conclude that the 'all to one' case is not valid above $\sim$3.5~eV. For event energies in the range $E_{g}$\textendash$2E_{g}$ (1.2\textendash2.4~eV in Si) only one electron-hole pair is allowed by energy conservation, forcing all charge yields to be insensitive to charge energy distribution and thus all models are identical. Above $2E_g$, impact ionization is possible, but the probability is strongly energy independent, and impact ionization only becomes appreciable when carriers exceed $\sim$2~eV of energy above gap.
    \item Around 4 -- 7.5~eV, our best-fit $\alpha$ rises quickly, indicating the distribution trends from an 'uniform split' ($\alpha=1$) to 'equal split' ($\alpha\rightarrow \infty)$.
    \item At $\sim$8~eV, where existing data stops, \eh\ appears to disfavour the charge yield predicted by the `all to one' model, while still in tension between the 'uniform split' and 'equal split' models. The range of best-fit $\alpha$ parameters implies that $P(E,E_r)$ is perhaps not captured effectively by a one parameter Beta distribution. 
    \item Above 8~eV \eh\ trends towards $\epsilon_{eh,\infty}$ with an oscillatory feature spanning between 14-20~eV. This is the signature of the finite value of $W$ regardless of the inclusion or value of \plasmon\ while the converse, $W \to \infty$ results in a smoothly rising function.
\end{enumerate}
The conclusions based on this empirical model agree with the energy distributions derived from density functional theory (DFT) in Ref. \cite{Wolf98}, where a local maximum in the quantum yield curve around 4.5~eV is attributed to a point of maximum energy sharing between electrons and holes, as shown in Fig.~\ref{fig:betadistribution}.

It is interesting to note that in the high energy-limit, where $W<<E_r$ should be indistinguishable from the all to one case, we nevertheless see a strong  modification to that parameter extreme on the resulting calculation of \eh. With this modification, the otherwise discrepant curve agrees with the best-fit and energy sharing models by 20~eV. The remaining uncertainty is therefore largely restricted to the 10\textendash20~eV energy range, where our best-fit model shows some non-linear behavior, and the different energy-sharing models are still not in agreement. Our exponential extrapolation largely splits the difference between the uniform and energy sharing distributions, and is only an approximate guess at the behavior here\footnote{The exponential extrapolation is just an ansatz that allows us to connect fits at 8~eV to data at 50~eV, and is a median model within the bounded behavior shown in Fig.~\ref{fig:ehfit}}. The full range of possible charge yields is bounded by the three energy sharing models, and does not include potential resonance features near e.g. the plasmon energy. On this point, no choice of \plasmon\ (varied from 14\textendash22~eV and $\to\infty$ effectively turning it off) modified the shape of the curves or $\epsilon_{eh,\infty}$ and $F_{\infty}$. This is likely attributed to the perfect down-conversion efficiency baked into our plasmon model. Suffice to say, experimental data in this energy regime is needed to refine the empirical model further.

\begin{figure}[ht]
	\centering
	\includegraphics[width=0.48\textwidth]{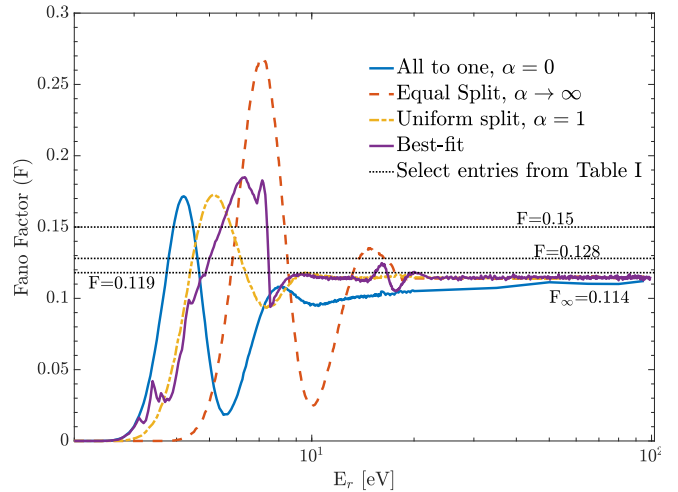}
	\caption{Computed Fano factor at 300~K for energy partitioning schemes discussed in text, along with curve (solid purple) from best-fit model. Dashed lines indicate values from literature as compiled in Table \ref{tab:measurements}.}
	\label{fig:fanolines}
\end{figure}

Fig.~\ref{fig:fanolines} demonstrates the non-constant behavior of the Fano factor at energies $<20$~eV. We once again note that the presented assumptions all match the observed asymptotic value $F_\infty$ by $\sim$100~eV but are all individually discrepant from measured behaviour below that, by a factor of 2$\times$ in some regimes. 

\subsection{Pair-creation probabilities}

\begin{figure}[t] 
	\centering
	\includegraphics[width=0.48\textwidth]{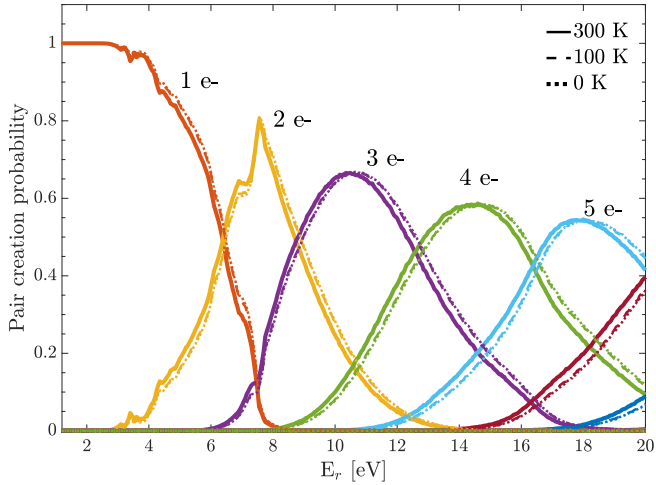}
	\caption{Pair-creation probability distributions for best-fit model at 0~K, 100~K and 300~K (former curves effectively overlap). These lines are to be interpreted as the probability to ionize the labeled number of charge pairs for a given deposited energy. These are \textit{not} PDFs in that only the sum of curves across a given point in energy is normalized to 1.}
	\label{fig:creationN}
\end{figure}

\begin{figure}[ht]
	\centering
	\includegraphics[width=0.48\textwidth]{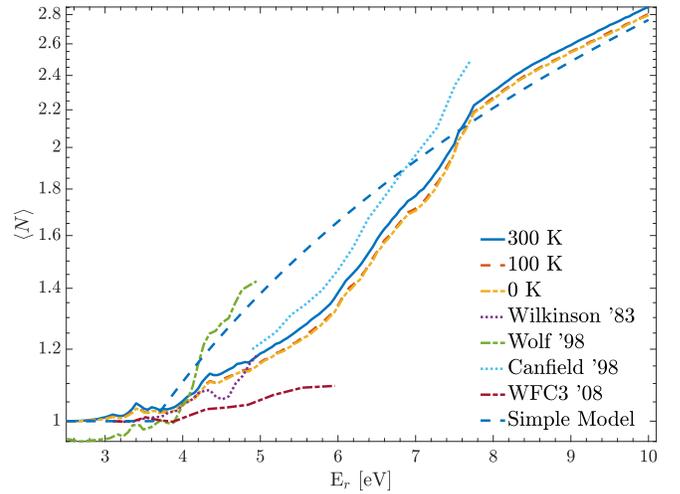}
	\includegraphics[width=0.48\textwidth]{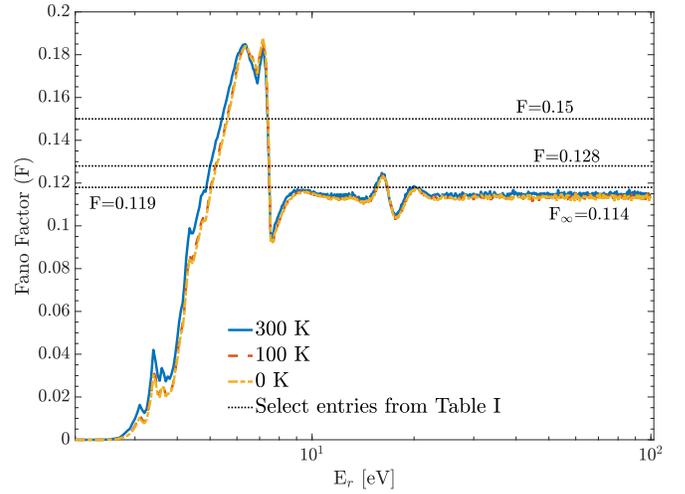}
	\caption{Top: The quantum yield, defined as the average number of charge pairs created at a given energy, for the best-fit model at 0K, 100K and 300K. Empirical data from Refs. \cite{Wolf98, wilkinson, Canfield_1998, borders2010wfc3} are provided as points of comparison. Bottom: Variation of the Fano factor \F for the best-fit model at 0K, 100K and 300K. While the asymptotic values, equivalent to those computed in Fig. \ref{fig:AEGdependence} (bottom right), are within 1\% of each other, there can be upwards of a $\sim$10\% difference at specific energy values.}
	\label{fig:bestFit}
\end{figure}

Tuning and validating the parameters of the Monte-Carlo model, as discussed in the prior section, enables us to produce the pair-creation probabilities seen in Fig. \ref{fig:creationN} for 3 temperature points around which many detectors operate at (see Appendix~\ref{app:supp}). The empirically derived nature of $p_{1,2,3}$ is clearly visible with smoother curves for $p_{>=4}$ where the model is reflective of the transition to an $\alpha=1$ regime. Fig.~\ref{fig:bestFit} illustrates the behavior of the best-fit quantum yield and Fano factor as a function of temperature compared to measurements from the literature, illustrating a significant dispersion of quantum yield measurements at a few eV and with our computed values lying centrally in this range. Data from Refs. \cite{Canfield_1998, wilkinson} comes from Si photodiodes at unspecified temperatures, assumed to be ambient, while Ref. \cite{borders2010wfc3} comes from averaged surface integration tests of the Hubble Wide Field Camera 3 CCDs at 224~K. 

We stress here the point that both \qy\ and \F\ are derived quantities, which are arguably only useful at ``high" energies where they are a shorthand for packaging the messy dynamics of ionization response with appeals to the central limit theorem. We argue that the probability of creating $n$ electron-hole pairs, $p_n$, is the preferential basis to understand charge yield by formulating these quantities in terms of ionization probability: $\langle N\rangle = \sum_{n=0}^{\infty} n p_n,\,\langle N^2\rangle = \sum_{n=0}^{\infty}n^2p_n$, and $F = \frac{\langle N^2 \rangle - \langle N \rangle^2}{\langle N \rangle}$ from which we recognize that the use of aggregate quantities, and exclusion of higher moments, informationally constrains both parameters. Stated more concretely, if $\sum_n\big[p_n(E)>0\big]>2$, as is true for most energies, then there are more terms than constraining equations and multiple solutions of $p_n$ would satisfy the same \qy\ and \F\ curves. 

However, due to the well behaved nature of \eh\ and \F\ for $E_r \gtrsim 50$~eV, where they are effectively flat, we can compute the exact Gaussian functional form, 
\begin{equation}
    p_n(E) = \frac{1}{\sqrt{2 \pi nF_{\infty}}}\rm{Exp}\bigg[{-\frac{1}{2}\bigg(\frac{n\epsilon_{eh,\infty} - E}{\sqrt{nF_{\infty}}\epsilon_{eh,\infty}}\bigg)^2}\bigg]
    \label{eq:pnexact}
\end{equation}
to infer $p_n$ in and beyond this region, sufficient for practical applications.

\subsection{Scientific Impact Example: DM Scattering}

\begin{figure*}[t]
	\centering
	\includegraphics[width=0.99\textwidth]{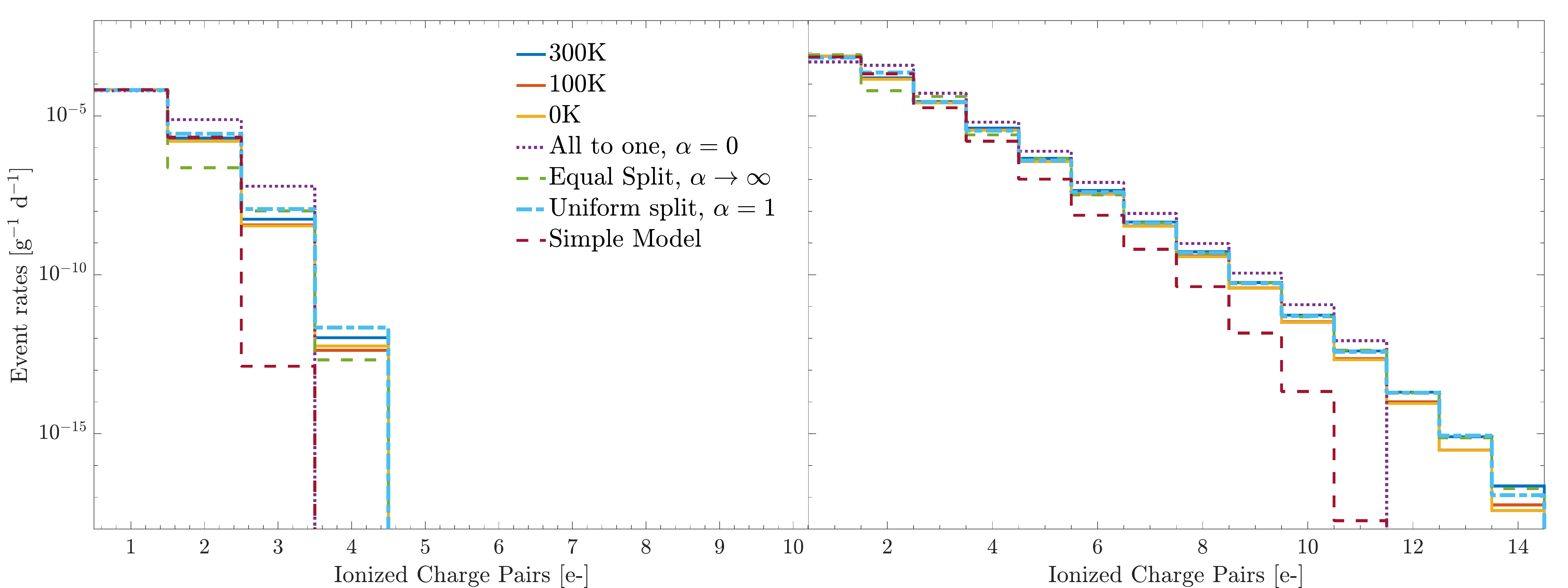}
	\caption{Dark matter electron scattering ionization spectra for the various ionization schemes presented in this paper, using a DM form factor $F_{DM} \propto 1$ and masses of 2~MeV (left) and 10~MeV (right).}
	\label{fig:mev_spectra}
\end{figure*}

\begin{figure*}[t]
	\centering
	\includegraphics[width=0.48\textwidth]{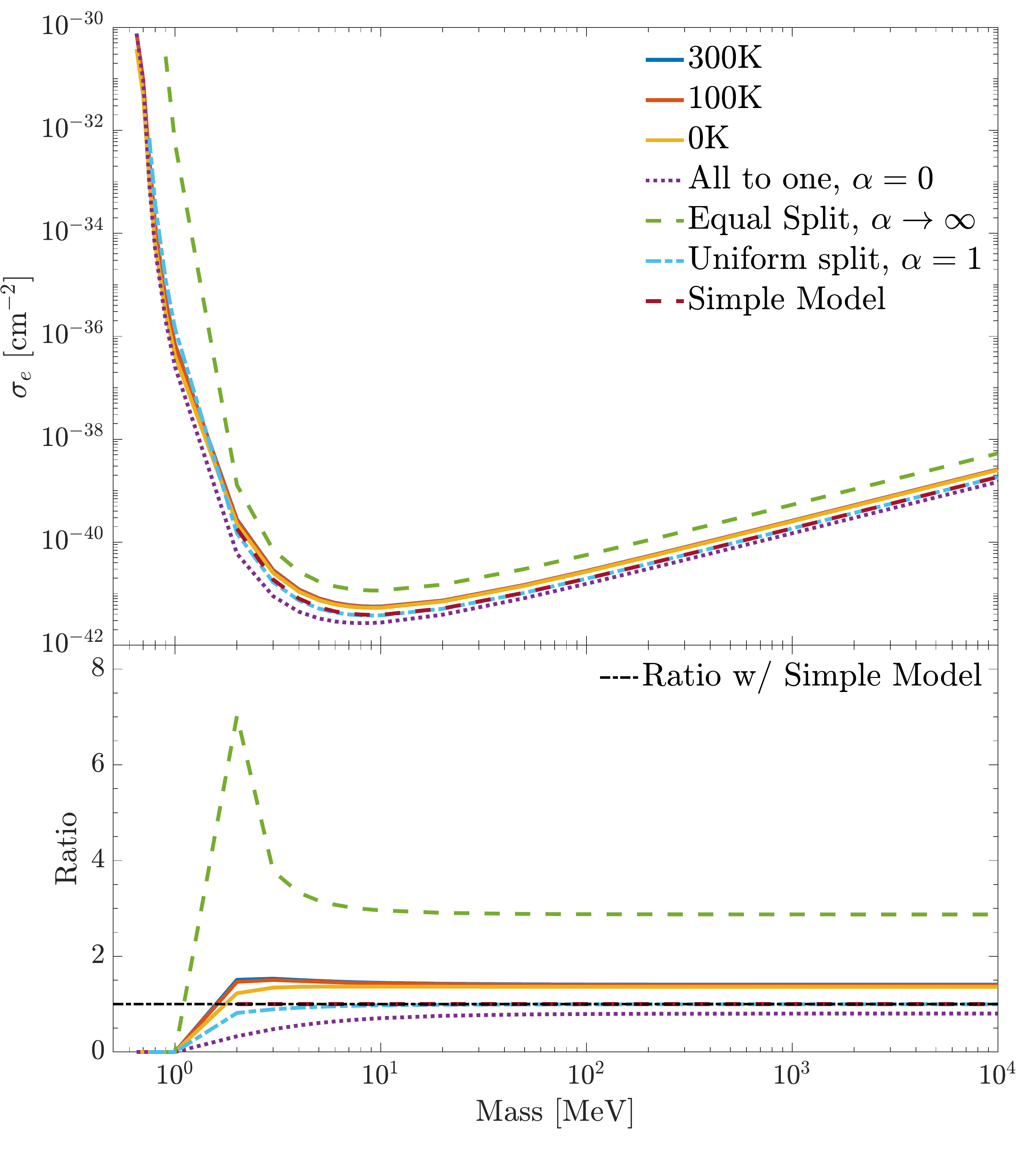}
	\includegraphics[width=0.48\textwidth]{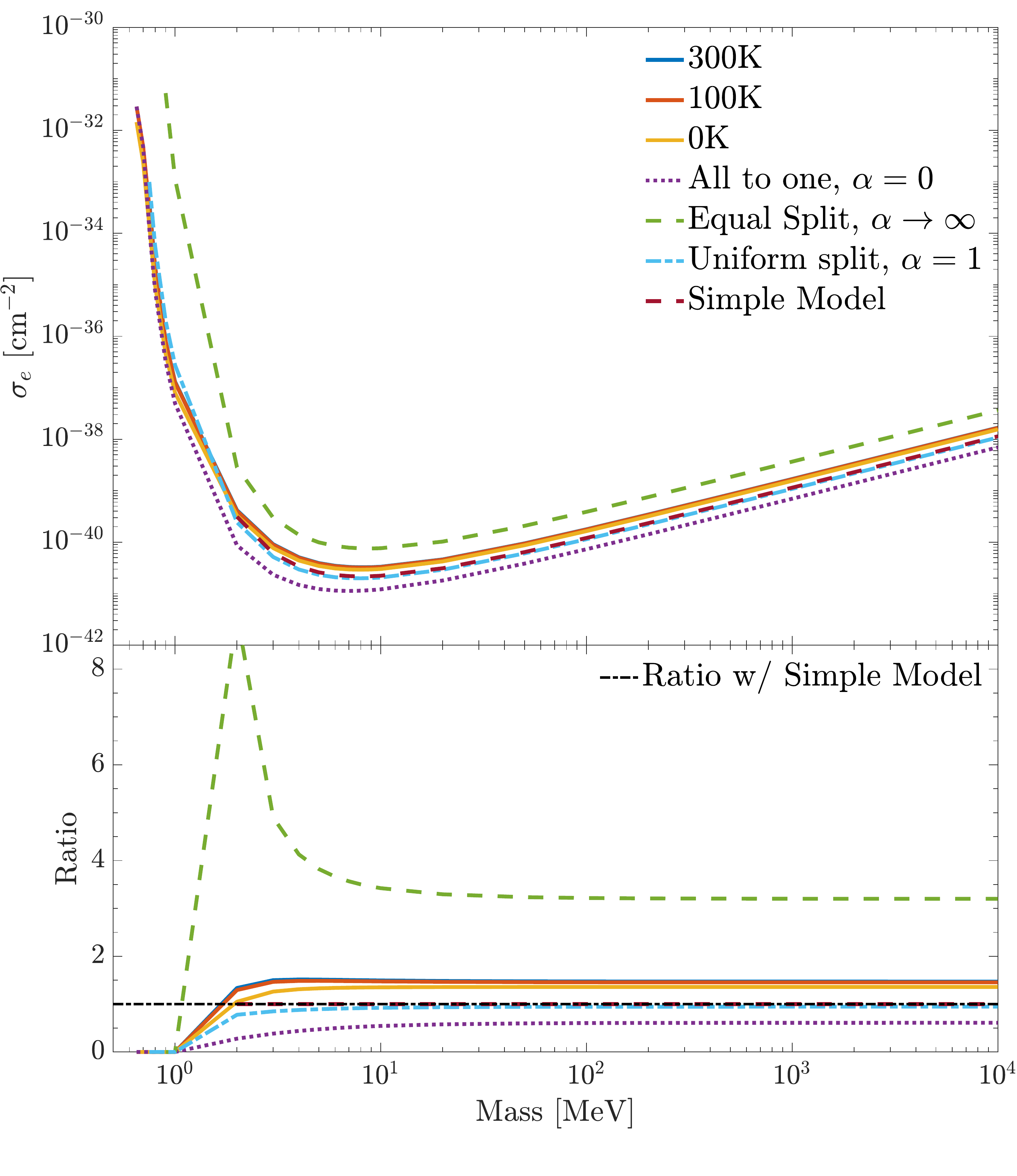}
	\caption{Dark matter electron scattering exclusion  curves  for  a  hypothetical  experiment assuming a 2 e$^-$ threshold and Poisson backgrounds for DM form factor $F_{\rm{DM}} \propto 1$ (left) and $F_{\rm{DM}} \propto q^{-2}$ (right). The ratios of the curves against the simple model are plotted below each, highlighting the around $50$\% discrepancy to computed limits accounting for silicon ionization micro-physics.}
	\label{fig:limits}
\end{figure*}

To investigate how low energy ionization statistics can potentially affect scientific results, we study the case of dark matter particles scattering off electrons. The bound nature of the electrons and crystalline band structure of the target requires us to follow the prescription of Ref. \cite{dm-electron} to compute scattering rates. Exploring the case of a 2~MeV and 10~MeV DM particle for form factors $F_{\rm{DM}} \propto q^{-k}$ $(k=0,2)$, after convolution with the various presented charge yield models, we see the ionization spectra represented in Fig. \ref{fig:mev_spectra}. The simple division of energies into bins of \eh\ dramatically underestimates the tails of the spectra by many orders of magnitude. Even with application of a charge yield model, 1 and 2 e$^{-}$ production rates are significantly different for varying $\alpha = 0 \to +\infty$. 
Translating these scattering spectra to a hypothetical direct detection experiment, under the assumption of a 2~e$^-$ threshold and Poisson background fluctuations, we can look at exclusion curves of electron-recoil dark matter scattering in a Si detector as presented in Fig. \ref{fig:limits}. These curves represent 90\% confidence level upper limits on the reference cross-section $\sigma_{e}$ for a 1 kg~year exposure with 0 observed events, as a function of DM mass. The effect of the charge yield modeling is pronounced at masses $<10$~MeV as seen by the lower panel ratio of the various models to the parsimonious simple model \textemdash\ revealing a difference of $\sim$50\% in limits when using a more accurate charge yield prescription. The case of $\alpha=0$ significantly underestimates limits, particularly at masses of 1\textendash5~MeV. Finally, the sensitivity of the experiment is different for the different cases, with lower mass thresholds varying from 0.5\textendash1~MeV, which is a purely model dependent effect and does not accurately reflect the true underlying physics.    

\section{Discussion}\label{sec:discussion}

We have provided a physically motivated charge yield model for a Si detector. By appealing to well-measured laboratory data, we have constructed an ionization response model valid between $\sim$1.2\textendash8 eV and further motivated an extension into the ``UV-gap" in which there are no current measurements. We have investigated how these probabilities vary with temperature and have explored the scientific impact of these models on a test case of electron-recoil dark matter. 

In contrast to the treatment in Ref.~\cite{dunford}, in which the Fano distribution is used to model charge yield down to the ionization threshold, we find that both \eh\ and \F\ are inadequate to accurately capture low-energy ionization yield. This is in large part due to the solid-state nature of Si; for processes close to the gap, where the phase space is restricted by the band-structure, we observe non-trivial departures from this simple two-parameter model. For processes at energies much larger than the band-gap, where carriers can be treated as free particles, we find that we recover the simple model.

While this model is the best current estimate of the behavior of charge yield due to electron-recoil processes in Si, we wish to highlight shortcomings of the model which further data will help to address. In particular, this model appeals to the plasmon (with energy $\sim$17~eV in Si) to explain the linearity at high-energy without any theoretical motivation for doing so. While the plasmon explanation provides a convenient heuristic, it is merely empirical, and is not predictive, as shown in Ref.~\cite{rothwarf}. In this work, we see no significant feature at the plasmon energy in any model. The parameter which has the largest impact on ionization yield is the width of the hole band, which introduces a non-linearity in the predicted yield for energies comparable to the hole band width, $\sim$12~eV, as shown in Fig.~\ref{fig:creationN}. Direct measurements of charge yield near the hole band edge and around the plasmon energy, within the UV gap, may yield more information about the relative importance of these processes at intermediate energies between the optical and soft X-ray data currently available.


In contrast, the general consensus ties the high-energy value of \eh\ only to the band-gap energy and impact ionization energies. The generic expression for \eh\ is \cite{roosbroeck,Klein,rothwarf,Canali72}
\begin{equation}
    \epsilon_{eh} = E_{g} + 2L\left[E_{i,e}+E_{i,h}\right] + E_r
\end{equation}
where $L$ is a factor which depends on the dispersion curve of the conduction and valence bands, $E_{i,e}$ and $E_{i,h}$ are the ionization thresholds for electrons and holes, and $E_r$ are phonon losses. Ref.~\cite{Canali72} shows that, for $E_{i,e}\sim E_{i,h} \propto E_{g}$, we get the formula
\begin{equation}
    \epsilon_{eh} = C \cdot E_{g} + E_r
\end{equation}
where $E_r$ takes on values from 0.25 to 1.2~eV, and $C$ is found to be $\sim$2.2 to 2.9~eV. Ref.~\cite{Klein} finds, using a broader range of materials, the parameters $C\sim2.8$ and $E_r\sim$0.5\textendash 1.0~eV. Studying materials with a fixed plasmon energy but varying gaps (such as the polytypes of SiC, with gaps ranging from 2.4 to 3.3~eV for a fixed plasmon energy\cite{griffin2020}) may help elucidate the role the plasmon plays, if any, in this down-conversion process.

Finally, we note that this model does not include the effects of inner shell electrons or any possible temperature dependence in the phenomenological constant $A$. The latter we expect to be a small effect, as the energy scales involved are higher than thermal energies at room temperature. The former have been noted to produce slight increases in the relative energy per pair (see e.g. Ref.~\cite{owens1996}), but only at the level of a few percent, and likely sub-dominant to statistical fluctuations for all but the most precise measurements. In addition, it is possible that charged particles and photons, which impart a different distribution of momenta to the electron-hole pairs, may require slightly different amounts of energy per subsequent pair created. Direct measurement of ionization yield by low energy electron recoils using Electron Energy Loss Spectroscopy (EELS) and an active target will allow for better characterization of the correspondence between electronic depositions from massive particles or photons.

\appendix
\section{Fano Factor Systematics}\label{app:fano}

A significant observation in this paper is that the Fano factors predicted by the model are lower than all of the existing measurements, which are inconsistent with each other. This can be accounted for by the one-sided systematic introduced into the measurement if finite charge collection efficiency (CCE) is not accounted for, or if other secondary processes can lead to impact ionization of additional charge in the crystal.

As an example of the systematic effect on measured Fano factor, we consider here the effect of finite charge collection. In this case, the probability of observing $n-k$ final charges given $n$ initial charges is \cite{kurinsky,Ponce_2020}
\begin{equation}
    P_{n-k} = \frac{n!}{k!(n-k)!}\eta^{n-k}(1-\eta)^k,
\end{equation}
where $\eta$ is the collection efficiency. For $\eta=1$, we find that $P_{n-k}=\delta(n-k)$ as expected. If we assume perfect charge resolution, we can calculate the measured mean ($n_{meas}$) and variance ($\sigma_{CCE}$) of the resulting charge distribution, which gives
\begin{align}
    n_{meas} &= \eta \cdot n, \\
    \sigma^2_{CCE} &= \eta \cdot n(1-\eta) = n_{meas}(1-\eta).
\end{align}
Given these moments, we thus get the measured Fano factor
\begin{align}
    F_{meas} &= \frac{\sigma^2_{meas}}{n_{meas}} \\
    &= \frac{\sigma^2_{fano}+\sigma^2_{CCE}}{n_{meas}} \\
    &\approx \frac{F\cdot n+\eta \cdot n(1-\eta)}{\eta \cdot n} \\
    &= \frac{F}{\eta}+(1-\eta)
\end{align}
where $F$ is the intrinsic Fano factor. Here the approximate sign comes from the fact that the CCE variance is slightly broadened due to the Fano factor as well; this approximation actually makes this estimate a lower bound on the measured Fano factor, but it is a small effect.

\begin{figure}[H]
    \centering
    \includegraphics[width=0.45\textwidth]{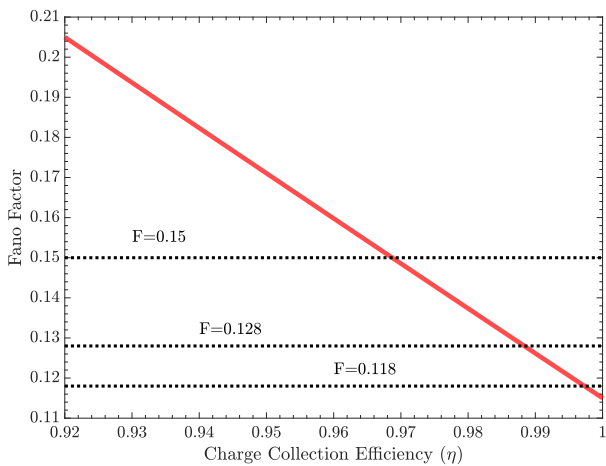}
    \caption{Measured Fano factor as a function of charge collection, assuming the true Fano factor lies in the range [0.112,0.115] predicted by our Monte Carlo results. The discrepancy between measurements can be accounted for by a small reduction in charge collection efficiency.}
    \label{fig:CCE}
\end{figure}

This is significant because $\eta$ does not fall out via averaging but presents as a measurement systematic. If we assume a true Fano factor of 0.115, for example, we only need charge collection efficiency to drop to 98\% in order to produce a measured Fano factor of 0.130, and a drop to better than 95\% gives a measured Fano factor of 0.160, the largest of the numbers we quote, as shown in Fig.~\ref{fig:CCE}. Absolute measures of charge collection are rarely made, and instead, voltage is converted to charge by assuming a known \eh\, given that these measurements have yet to be done with single charge-resolving detectors. Single charge resolving detectors, which can more easily characterize CCE by observing partial collection of single electron-hole pairs (see e.g. Ref.~\cite{Ponce_2020}), promise to significantly reduce these systematics, and should be able to produce much more accurate measurements of intrinsic Fano factor. In particular, the recent measurement of Ref.~\cite{rodrigues2020} with a skipper CCD at 6~keV is within error of our model in both $\epsilon_{eh}$ and $F$ and demonstrated CCE much better than 90\%. A more in-depth discussion of Fano factor systematics can be found in e.g. Ref.~\cite{eberhardt1970fano}. 

\section{Supplementary Material}\label{app:supp}

We provide three tab-delimited flat files (\textit{p0K.dat}, \textit{p100K.dat}, and \textit{p300K.dat}) containing the quantity $p_n(E)$ computed in this paper at the three different reference temperatures. The first column of each file is energy $E_{g} \leq E \leq 50$~eV, while consecutive columns are $p_n$ for $n=[1..20]$. The rows entries are normalized to one and are straightforwardly interpreted as probabilities. \\

\section*{Acknowledgements}
We would like to thank, in no particular order, Dan Baxter, Alvaro Chavarria, Rouven Essig, Juan Estrada, Yonatan Kahn, Matt Pyle, Alan Robinson, Kyle Sundqvist, Javier Tiffenberg, Belina von Krosigk, and Matt Wilson for useful discussions related to the model described in this work. We would also like to thank Lauren Hsu for feedback on an early draft of the paper. We thank the Gordon and Betty Moore Foundation and the American Physical Society for the support of the ``New Directions in Light Dark Matter'' workshop, at which the idea for this paper was conceived and fruitful discussion occurred. We acknowledge financial support from the Kavli Institute for Cosmological Physics at the The University of Chicago through an endowment from the Kavli foundation, and the National Science Foundation through Grant No. NSF PHY-1806974. This document was prepared by N.K. using the resources of the Fermi National Accelerator Laboratory (Fermilab), a U.S. Department of Energy, Office of Science, HEP User Facility. Fermilab is managed by Fermi Research Alliance, LLC (FRA), acting under Contract No. DE-AC02-07CH11359.

\bibliography{refs.bib}

\end{document}